\pdfoutput=1

\documentclass[11pt]{article}
\usepackage[table,dvipsnames]{xcolor}

\usepackage[final]{acl}

\usepackage{times}
\usepackage{latexsym}

\usepackage[T1]{fontenc}

\usepackage[utf8]{inputenc}

\usepackage{microtype}

\usepackage{inconsolata}

\usepackage{xspace}
\usepackage{graphicx}
\usepackage{multirow}
\usepackage{amsfonts}
\usepackage{amsmath}
\usepackage{subfigure}
\usepackage{comment}
\usepackage{booktabs}

\usepackage{amssymb}
\usepackage{pifont}
\usepackage{makecell}
\usepackage{subscript}
\usepackage{hyperref}
\usepackage{listings}
\usepackage{array}
\usepackage{tabularx}

\author{
    Dohyeon Lee\textsuperscript{\rm 1}, 
    Yeonseok Jeong\textsuperscript{\rm 2},
    Seung-won Hwang\textsuperscript{\rm 1 \rm 2}\thanks{~~Corresponding Authors}
    \\
    Computer Science and Engineering, Seoul National University\textsuperscript{\rm 1},
    \\
    Interdisciplinary Program in Artificial Intelligence, Seoul National University\textsuperscript{\rm 2}
    \\
    \texttt{\{waylight3,  jys3136, seungwonh\}@snu.ac.kr} \\
}

\newcommand{\ours}{SMR\xspace}

\newcommand{\oursfullhighlight}{\underline{\textbf{S}}tate
\underline{\textbf{M}}achine
\underline{\textbf{R}}easoning\xspace}

\title{From Token to Action: State Machine Reasoning to Mitigate Overthinking in Information Retrieval}

\begin{document}
\maketitle

\begin{abstract}
Chain-of-Thought (CoT) prompting enables complex reasoning in large language models (LLMs), including applications in information retrieval (IR).
However, it often leads to overthinking, where models produce excessively long and semantically redundant traces with little or no benefit.
We identify two key challenges in IR: redundant trajectories that revisit similar states and misguided reasoning that diverges from user intent.
To address these, we propose \oursfullhighlight (\ours), a transition-based reasoning framework composed of discrete actions (\textsc{Refine}, \textsc{Rerank}, \textsc{Stop}) that support early stopping and fine-grained control.
Experiments on the BEIR and BRIGHT benchmarks show that \ours improves retrieval performance (nDCG@10) by 3.4\% while reducing token usage by 74.4\%.
It generalizes across LLMs and retrievers without requiring task-specific tuning, offering a practical alternative to conventional CoT reasoning.\footnote{
The code and details are available at \url{https://github.com/ldilab/SMR}.}
\end{abstract}

\section{Introduction}
\label{sec:intro}

\begin{figure}[t]
    \centering
    \includegraphics[width=1.0\columnwidth]{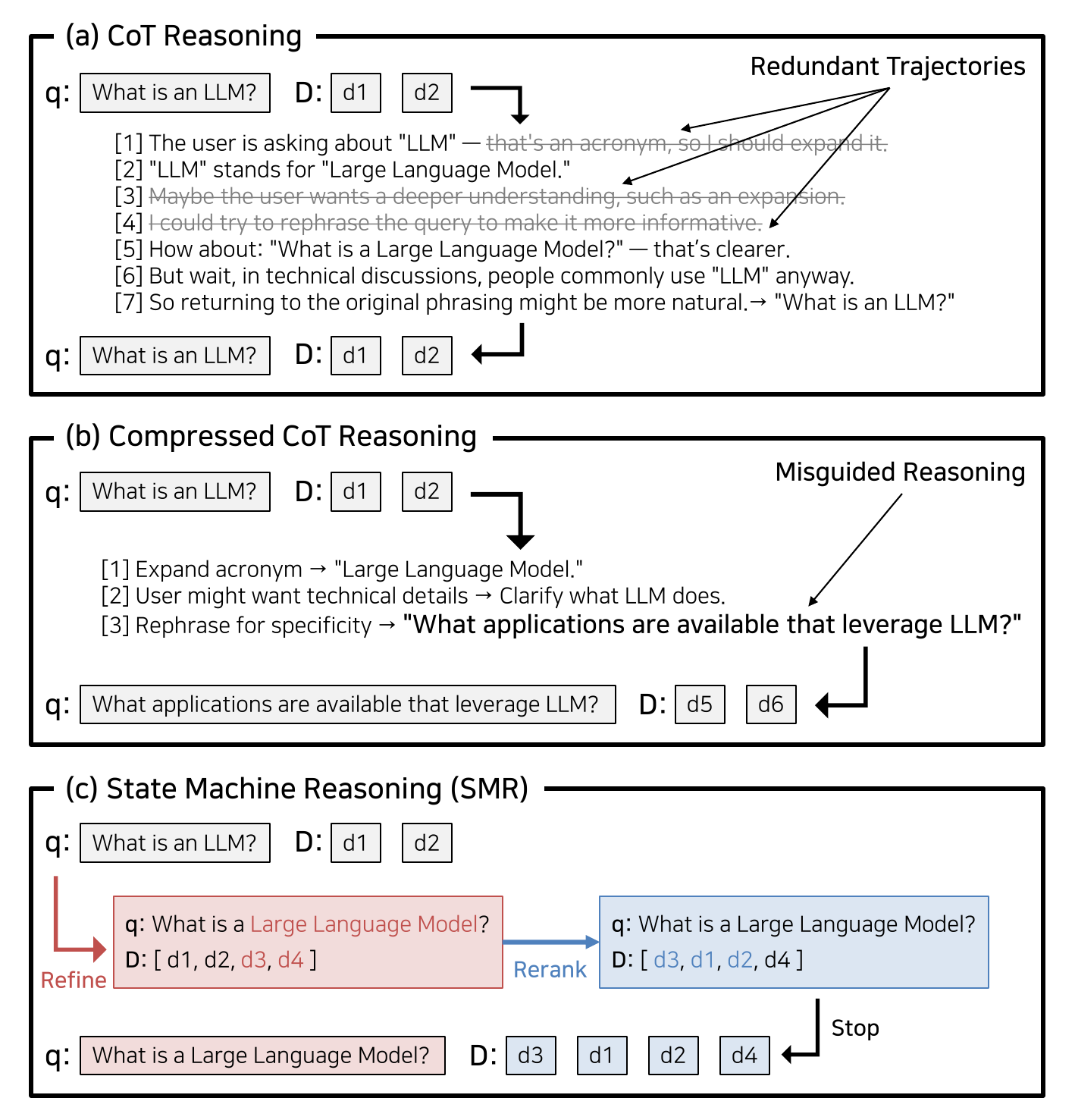}
    \caption{Conceptual comparison between standard CoT reasoning, compressed CoT reasoning, and our proposed \ours framework. (a) Standard CoT performs token-level generation in a single forward pass, often leading to redundant intermediate states. (b) Compressed CoT shortens trajectories via reinforcement learning, at the risk of misaligned outputs. (c) \ours decomposes reasoning into structured state transitions $(q, D)$ guided by IR-specific actions.}
    \label{fig:intro}
\end{figure}

Chain-of-Thought (CoT) prompting~\cite{wei2022chain} has emerged as a powerful paradigm for enhancing large language model (LLM) reasoning across complex tasks~\cite{guo2025deepseek,qwq32b} such as math and planning. This paper focuses on extending CoT-based reasoning to information retrieval (IR)~\cite{li2025search,shao2025reasonir}, to better align retrieval with user intent for improved query rewriting~\cite{ma2023query} and document reranking~\cite{sun2023chatgpt}.

Despite its strengths, CoT prompting often suffers from \emph{overthinking}~\cite{chen2024not,liu2024mind,fan2025missing}—the generation of semantically redundant or incorrect reasoning steps that offer no benefit and may even be detrimental. We identify two key IR-specific challenges arising from this inefficiency that remain unaddressed by current methods.

\paragraph{Redundant Reasoning}
Standard CoT prompting generates reasoning at the token level, often revisiting semantically equivalent steps and continuing unnecessarily even after reaching the answer.
As illustrated in \autoref{fig:intro}(a), CoT-based query rewriting frequently produces redundant paraphrases that fail to introduce new evidence for retrieval, e.g., text shown in grey in the figure, which only inflate with no gain in inference.
A useful evidence that is missing would be a passage mentioning LLM in a full name, to retrieve a higher recall results including d3 and d4.

\paragraph{Misguided Reasoning}
In an attempt to compress CoT trajectories using reinforcement learning (RL) to shorten reasoning paths, as in O1-Pruner~\cite{luo2025o1}, reasoning may misalign to user intent in open-ended domain~\cite{li2025whenthinkingfails}.
As illustrated in the bold text of \autoref{fig:intro}(b), such compression can result in syntactically concise yet semantically misaligned queries, e.g., retrieving LLM ``applications,'' drifting from an original intention of its ``definition''.
This misalignment causes the retrieval of irrelevant documents such as d5 and d6.
Furthermore, compression requires task-specific training and reward engineering, and thus limits generalization.

To address these challenges, we propose \oursfullhighlight (\ours), a transition-based reasoning framework based on structured state transitions that avoids token-level generation.
Instead of relying on autoregressive decoding or compressing reasoning trajectories, \ours formulates reasoning as transitions between intermediate representation states.
Inspired by decision-theoretic frameworks~\cite{liu2023state}, \ours represents each step as a transition between states $(q, D)$, where $q$ denotes the current query, and $D$ a ranked list of retrieved documents.
\autoref{fig:intro}(c) illustrates how \ours reaches the goal.
Our policy model first selects the \textsc{Refine} action to expand the acronym ``LLM,'' followed by the \textsc{Rerank} action to reorder the newly found document d3, which is the most relevant.
This policy model terminates when next step produces no gains, by selecting
 \textsc{Stop} action.
This mechanism ensures that each state transition yields a incremental gain and preserves the evolving context through structured state representations.
This formulation brings two key advantages.

\paragraph{Token Efficiency}
Our method avoids redundant reasoning and achieves greater token efficiency by operating over explicitly defined states.
Each step updates a structured state, allowing the system to detect when it revisits an equivalent state.
This contrasts with token-level generation, which lacks a mechanism to recognize semantic repetition.

\paragraph{Action Effectiveness}
By grounding each reasoning step in IR-relevant operations, \ours enables retrieval systems to make improvements through two actions: \textsc{Refine} for query rewriting and \textsc{Rerank} for document reranking.
These actions enable the system to reissue queries when current results are insufficient, or reorder documents when initial rankings are suboptimal.
This design supports incremental retrieval improvements and offers precise control over which component of the pipeline to adjust at each step.
In contrast, token-level generation lacks explicit validation of improvement at each step, can resulting in outputs that diverge from user intent.

We validate \ours on two benchmarks: BEIR~\cite{thakur2021beir}, a widely-used benchmark for standard IR evaluation, and BRIGHT~\cite{su2024bright}, which emphasizes reasoning-intensive retrieval.
\ours achieves up to \textbf{3.4\%} improvement in nDCG@10 while reducing token usage by \textbf{74.7\%} on the BRIGHT benchmark, consistently outperforming prior baselines across both retrieval quality and efficiency metrics.
\ours demonstrates robustness across  sparse/dense retrievers and LLMs with and without reasoning.
These results validate the practicality and generalizability of our approach, suggesting that \ours can be seamlessly integrated into a wide range of retrieval systems without requiring model-specific tuning.

\section{Related Work}
\label{sec:related}

\subsection{LLM-based Reasoning for IR}
LLMs have recently been used to interleave reasoning with retrieval actions, enhancing retrieval quality via query understanding and iterative refinement.
While early neural retrievers relied on static query-document matching~\cite{karpukhin2020dense,nogueira2019passage}, later approaches incorporated reasoning modules into the retrieval loop.
These methods typically operate at the token level, relying on long reasoning traces that are not always efficient or robust.
ReAct~\cite{yao2023react} pioneered interleaved reasoning and acting for multi-hop QA, inspiring follow-up work in CoT-based query rewriting~\cite{li2025search} and document reranking~\cite{weller2025rank1,zhuang2025rank}.
While ReAct interleaves reasoning and tool use, it lacks task-specific structure and often relies on verbose traces, limiting its efficiency in retrieval-focused tasks.

\subsection{Overthinking in CoT Reasoning}
Despite their effectiveness, CoT-based models often suffer from \emph{overthinking}, unnecessarily long reasoning sequences that offer no additional semantic gain~\cite{chen2024not,liu2024mind,fan2025missing}.
In the context of IR, this manifests as two distinct problems:
(1) \emph{Redundant trajectories}, where semantically similar steps are revisited repeatedly, and
(2) \emph{Misguided reasoning}, where shortened traces drift away from user intent or task objectives.
These issues increase inference cost and degrade alignment with the user’s retrieval goal.

\subsection{Overthinking Mitigation}
Recent work~\cite{sui2025stop, qu2025survey, wang2025harnessing} has investigated approaches to mitigate overthinking by compressing CoT traces or learning to generate shorter ones.
O1-Pruner~\cite{luo2025o1} applies RL to shorten reasoning while preserving output correctness.
Other methods use value estimation~\cite{shridhar2023distilling} or supervised distillation to train models to prefer concise traces.

\subsection{Our Distinctions}
Compared to LLM-based reasoning like ReAct (Section 2.1), our controlled action space enables a principled transition-based control against overthinking, without incurring task-specific training or reward engineering of existing work (Section 2.2).
We are the first to generalize state machine, beyond math and code~\cite{liu2023state} where verification is straightforward, to retrieval.
We show \ours is tuning-free, generalizable, and performs consistently across retrievers and LLMs.

\section{Proposed Method}
\label{sec:method}
Our framework can be viewed as a Markov Decision Process (MDP) with a discrete action space and structured state representation.
While we do not explicitly learn a value function, our design is inspired by decision-theoretic frameworks~\cite{liu2023state}, where each reasoning step corresponds to a transition between abstract states.
Based on this formulation, we recast reasoning as transitions over IR-specific states and actions, rather than as token-level generation.

\begin{figure*}[t]
    \centering
    \includegraphics[width=1.0\linewidth]{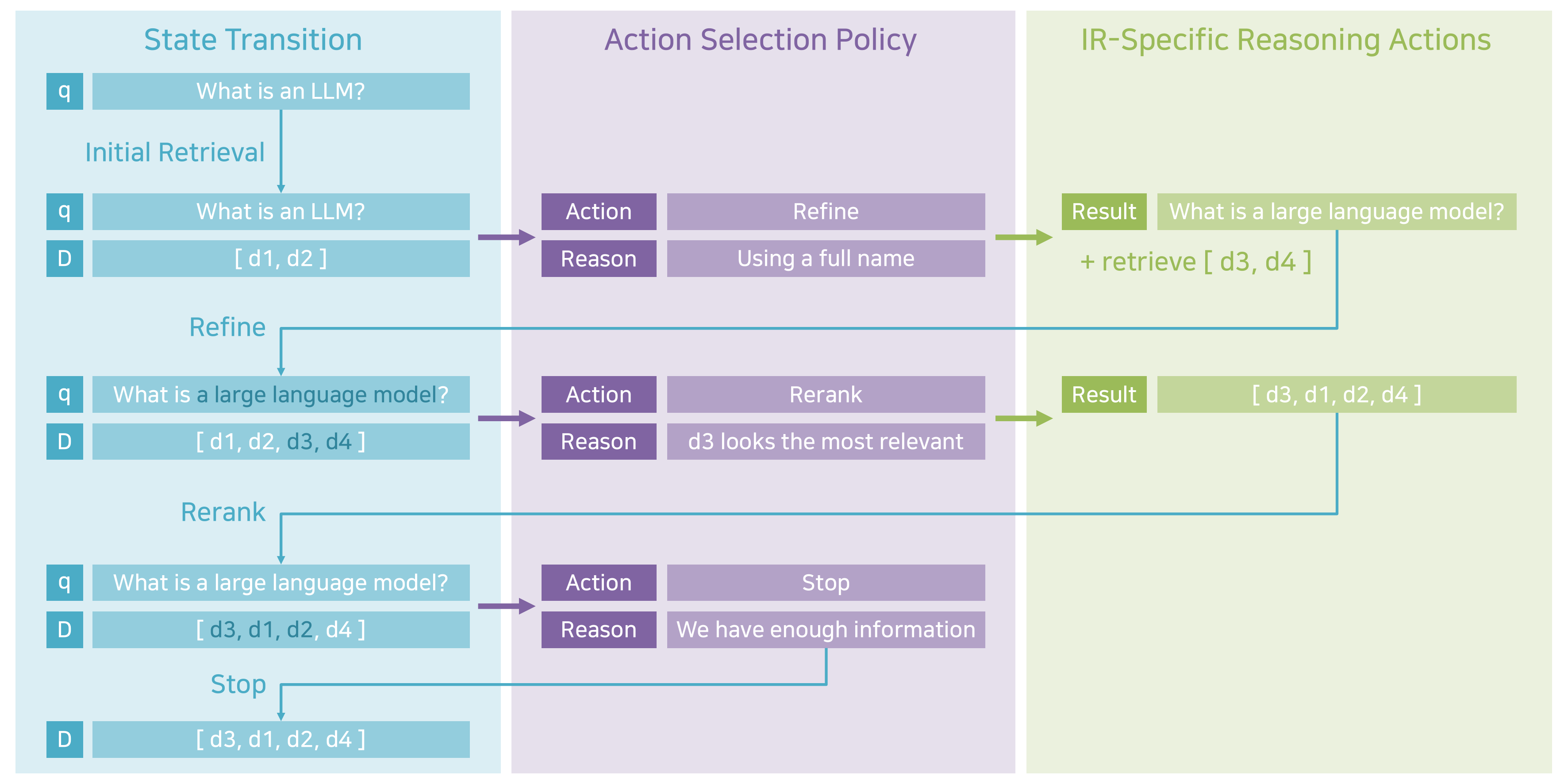}
    \caption{Illustration of our proposed reasoning framework (\ours). Beginning from an initial query and its retrieved documents, \ours transitions through structured states via three actions: \textsc{Refine} (query rewriting), \textsc{Rerank} (document reordering), and \textsc{Stop} (termination). At each step, the LLM selects an action with justification, and the document list is updated accordingly.}
    \label{fig:ours}
\end{figure*}

\subsection{Structured State Representation}
\label{sec:state}
To address the first challenge (redundant trajectories), we propose representing the reasoning state as a structured tuple $(q_t, D_t)$ at each step $t$, where $q_t$ denotes the current query and $D_t$ is a ranked list of top-$k$ retrieved documents:

\begin{equation}
s_t = (q_t, D_t), \quad D_t = \{d_1, d_2, \dots, d_k\}.
\end{equation}

This representation serves as the foundation of our transition-based reasoning by explicitly encoding the query and its retrieved documents.
Specifically, the query $q$ preserves the user’s intent, while the document list $D$ encapsulates the retrieval context that guides subsequent reasoning steps.

The initial state $s_0 = (q_0, D_0)$ is constructed using the user-issued query $q_0$ and the corresponding retrieved documents $D_0$ obtained from a static retriever.
Subsequent reasoning steps update either the query or the document list, producing a trajectory of structured states.
This formulation guides reasoning via IR-specific actions while maintaining alignment with retrieval quality and intent.

\paragraph{Stop Decision}
To avoid redundant reasoning and mitigate overthinking, \ours employs a stopping mechanism that detects when the system has reached an equivalent state.
Instead of explicitly comparing the current state $s_t = (q_t, D_t)$ with all previously visited states $\{s_0, s_1, \dots, s_{t-1}\}$, we assess whether the current state provides any meaningful gain.
Specifically, if the retrieved documents are identical and the query remains unchanged from the previous state $s_{t-1}$, then $s_t$ is treated as equivalent to $s_{t-1}$.
This ensures that state transitions reflect incremental improvements rather than redundant cycles.
Crucially, this mechanism allows the model to determine whether it has gathered sufficient information to stop reasoning, a capability that conventional CoT-based methods lack.
This decision is primarily made by the policy model, which selects the \textsc{stop} action when it predicts that further reasoning would yield no meaningful change.

\subsection{IR-Specific Reasoning Actions}
\label{sec:action}
To address the second challenge (misguided reasoning), we replace token-level decoding with a set of controllable reasoning actions.
This action space allows the system to updates the current reasoning state based on its retrieval context, rather than relying on implicit generation dynamics that may drift from user intent.

Our action space $\mathcal{A}$ consists of three discrete actions:
\begin{equation}
\mathcal{A} = \{ \textsc{Refine}, \textsc{Rerank}, \textsc{Stop} \}.
\end{equation}
Each action modifies the reasoning state to improve retrieval quality or terminate the reasoning process.

\paragraph{Refine}
The \textsc{Refine} action updates the query to better reflect the user's information need, guided by the current retrieval context:
\begin{equation}
q_{t+1} = \text{LLM}_{\textsc{refine}}(q_t, D_t).
\end{equation}
This enables the model to perform query rewriting or expansion, conditioned on evidence from the retrieved documents.

Following each \textsc{Refine} action, we invoke the retriever with the updated query $q_{t+1}$ to obtain new candidate documents.
If any of the retrieved documents are not already present in $D_t$, they are appended to the end of the current list.
This augmentation strategy ensures meaningful evolution of the retrieval state without discarding the existing context.
For further details on the \textsc{Refine} action, refer to Appendix~\ref{appendix:detail_refine}.

\paragraph{Rerank}
The \textsc{Rerank} action adjusts the ordering of the document list without modifying the query:
\begin{equation}
D_{t+1} = \text{LLM}_{\textsc{rerank}}(q_t, D_t).
\end{equation}
It refines the document ranking when the initial retrieval is imperfect, allowing better relevance estimation while keeping the query fixed.

To address potential hallucinations during reranking, we impose structural constraints on the output.
Specifically, if the reranked list contains documents that were not present in the original $D_t$, we discard those entries and exclude them from the updated state.
This prevents spurious hallucinated documents from contaminating the reasoning trajectory.
Conversely, if the reranked list omits any documents from the original set, we re-append the missing items to the end of the list in their original order.
This ensures that no context is inadvertently lost during reranking, preserving the integrity of the retrieval state across transitions.

\paragraph{Stop}
The \textsc{Stop} action terminates the reasoning process, returning the current state $s_t = (q_t, D_t)$ as the final output.
This allows the system to avoid unnecessary steps once sufficient retrieval quality is achieved, promoting token efficiency and preventing semantic drift.

In addition to semantic equivalence, we also impose a hard cap on the total number of reasoning steps to control inference cost.
Specifically, we set a maximum number of transitions, typically \texttt{max\_steps} = 16, beyond which the system automatically issues a \textsc{Stop} action.
This provides an upper bound on computation and ensures robustness in deployment scenarios with limited resources.
When compute budget permits, increasing \texttt{max\_steps} can yield improved retrieval performance by allowing deeper reasoning.

By decoupling reasoning into these interpretable transitions, our framework enables targeted improvements at each stage, supports early stopping, and remains robust across diverse retrieval scenarios without task-specific training.

\subsection{Action Selection Policy}
\label{sec:policy}
At each reasoning step, the system selects one of the three available actions based on the current state $(q_t, D_t)$.
Instead of relying on a fixed heuristic or trained controller, we adopt a prompt-based strategy where an LLM itself determines the next action.
In this setup, the LLM serves as a judge that evaluates the current reasoning context and chooses the most appropriate next step.

At each step, the system selects the next action based on the current state using an instruction-following LLM guided by a structured prompt.
The prompt describes the agent’s role as a decision-maker responsible for improving the quality of retrieved results, and presents the current query and its associated documents in a structured format.
The LLM is prompted to choose exactly one of the three actions.
The full prompt format is provided in Appendix~\ref{appendix:prompt_policy}.

Rather than relying on learned scoring or implicit decoding dynamics, the action decision is made through rule-grounded prompting that embeds heuristics into the LLM's instruction space.
For example, the LLM is encouraged to choose \textsc{Refine} when the query appears vague or the results are unsatisfactory, and to prefer \textsc{Stop} only when confident that no further improvement is possible.
This design allows the decision process to remain transparent, easily modifiable, and robust to misalignment.

While the rationale is not used by the system for downstream decisions, we retain it for transparency and interpretability.
This enables human inspection of the model’s behavior and helps ensure that the policy aligns with the intended design.
Remaining details of \ours are provided in Appendix~\ref{appendix:detail_policy}.

\section{Results and Analysis}
\label{sec:results}

\subsection{Experimental Setup}

\paragraph{Datasets}
We evaluate our method on two retrieval benchmarks with differing reasoning complexity.
\textbf{BRIGHT}~\cite{su2024bright} is a recent benchmark designed to evaluate reasoning-intensive retrieval scenarios.
It consists of 12 subsets spanning diverse domains such as StackExchange forums, coding problems, and STEM question-answering tasks, where queries often require domain-specific reasoning to identify relevant passages.
This serves as our primary benchmark, enabling us to assess how well different methods handle complex reasoning demands.
For completeness, we also evaluate on \textbf{BEIR}~\cite{thakur2021beir}, a widely used benchmark comprising diverse domains (e.g., factoid QA, biomedical search, and entity-centric retrieval).
BEIR primarily measures general retrieval performance and helps validate the effectiveness of our method in standard IR settings.

\paragraph{Evaluation Metrics}
We report \textbf{nDCG@10} as the primary evaluation metric, following standard IR practice.
Additional ranking metrics such as MAP and Recall are reported in the Appendix~\ref{appendix:result_bright_bm25_metric} for completeness.

\paragraph{Baselines}
We compare \ours against the following baselines:
\begin{itemize}
    \item \textbf{Retrievers}: BM25~\cite{robertson2009bm25} is a traditional sparse retriever widely used in IR. ReasonIR~\cite{shao2025reasonir} is a state-of-the-art dense retriever specifically trained for general reasoning tasks, outperforming other strong retrievers such as Contriever~\cite{izacard2021unsupervised} and RankLLaMA~\cite{ma2024fine}.
    \item \textbf{Standard CoT}: Rank1~\cite{weller2025rank1} and Rank-R1~\cite{zhuang2025rank} are CoT-based reasoning models that represent the state-of-the-art among fine-tuned and RL-trained approaches. We select these models over ReAct~\cite{yao2023react} as they are specifically adapted for IR and represent the leading CoT-based reasoning in this domain.
    \item \textbf{Compressed CoT}: O1-Pruner~\cite{luo2025o1} is an RL-based method for compressing reasoning trajectories while preserving answer quality.
\end{itemize}
To ensure a fair comparison across LLMs with differing capabilities, we experiment with both particular non-reasoning and reasoning LLMs (Qwen2.5-32B, QwQ-32B).
All baselines follow public implementations provided in official repositories, with necessary adaptations for our experimental protocol.

\paragraph{Implementation Details}
See Appendix~\ref{appendix:implementation_details}.

\subsection{Analysis}
To further validate the effectiveness of \ours, we assess whether our goals have been achieved through the following three research questions.

\begin{table*}[t]
\centering
\scalebox{0.78}{
\def\arraystretch{1.3}
\begin{tabular}{lcccccccccccc|c}
\hline
\multicolumn{1}{l|}{}                    &  Bio & Earth & Econ &  Psy &  Rob & Stack &  Sus & Leet & Pony & AoPS & TheoQ & TheoT &  \textbf{Avg} \\
\hline
\rowcolor{gray!20}
\multicolumn{14}{l}{\textit{\textbf{Retriever}}}                                                                                                 \\
\hline
\multicolumn{1}{l|}{BM25}                & 18.9 &  27.2 & 14.9 & 12.5 & 13.6 &  18.4 & 15.0 & 24.4 &  7.9 &  6.2 &  10.4 & 4.9   &          14.5 \\
\hline
\rowcolor{gray!20}
\multicolumn{14}{l}{\textit{\textbf{CoT Reasoning}}}                                                                                             \\
\hline
\multicolumn{1}{l|}{Rank1 (32B)}         & 22.1 &  31.7 & 14.6 & 15.7 & 15.8 &  17.7 & 20.3 & 22.9 &  \textbf{9.7} &  5.9 &  11.8 & 9.3   &          16.5 \\
\multicolumn{1}{l|}{Rank-R1 (14B)}       & 23.6 & 33.5  & 16.8 & 15.4 & 18.8 &  18.4 & 20.3 & 24.9 &  9.1 &  \textbf{6.9} &  12.1 & 8.7   &          17.4 \\
\hline
\rowcolor{gray!20}
\multicolumn{14}{l}{\textit{\textbf{Compressed CoT Reasoning}}}                                                                                  \\
\hline
\multicolumn{1}{l|}{O1-Pruner (32B)}     & 23.6 &  31.9 & 17.7 & 18.0 & 17.2 & 20.0  & 19.8 & 24.2 & 8.4  & 6.6  & 10.7  & 7.0   &          17.1 \\
\hline
\rowcolor{gray!20}
\multicolumn{14}{l}{\textit{\textbf{Ours}}}                                                                                                      \\
\hline
\multicolumn{1}{l|}{\ours (Qwen2.5-32B)} & 28.7          & \textbf{34.9} & \textbf{20.4} & 20.7          & \textbf{20.9} & 20.8          & 19.2          & 22.1          & 6.3          & 6.3 & \textbf{18.0} & \textbf{20.3} & \textbf{19.9} \\
\multicolumn{1}{l|}{\ours (QwQ-32B)}     & \textbf{28.8} & 34.4          & 19.8          & \textbf{21.0} & 19.8          & \textbf{21.1} & \textbf{21.4} & \textbf{25.5} & 7.5 & 5.7          & 16.9          & 17.2          & \textbf{19.9} \\
\hline
\end{tabular}
}
\caption{Retrieval performance (nDCG@10) on BRIGHT benchmark using \textbf{sparse retriever (BM25)} as the underlying retriever. All methods differ only in their reasoning strategy. Best scores per dataset are bolded.}
\label{tab:result_bright_bm25}
\end{table*}

\begin{table*}[t]
\centering
\scalebox{0.78}{
\def\arraystretch{1.3}
\begin{tabular}{lcccccccccccc|c}
\hline
\multicolumn{1}{l|}{}                    &  Bio & Earth & Econ &  Psy &  Rob & Stack &  Sus & Leet & Pony & AoPS & TheoQ & TheoT &  \textbf{Avg} \\
\hline
\rowcolor{gray!20}
\multicolumn{14}{l}{\textit{\textbf{Retriever}}}                                                                                                 \\
\hline
\multicolumn{1}{l|}{ReasonIR}            & 26.3 & 31.5 & 23.3 & 30.3 & 17.8 & 24.0 & 20.6 & \textbf{35.0} & 10.3 & 14.3 & 31.6 & 27.2 & 24.4 \\
\hline
\rowcolor{gray!20}
\multicolumn{14}{l}{\textit{\textbf{CoT Reasoning}}}                                                                                             \\
\hline
\multicolumn{1}{l|}{Rank1 (32B)}         & 32.0 &  30.0 & 22.3 & 31.1 & 16.7 & 25.8 & 22.4 & 31.4 & \textbf{15.5} & 12.1 &  27.8 & 26.3  &          24.5 \\
\multicolumn{1}{l|}{Rank-R1 (14B)}       & 33.0 &  34.1 & 24.2 & 33.5 & 20.2 & 25.4 & 22.8 & 33.5 & 14.1 & 10.7 &  30.3 & 27.8  &          25.8 \\
\hline
\rowcolor{gray!20}
\multicolumn{14}{l}{\textit{\textbf{Compressed CoT Reasoning}}}                                                                                  \\
\hline
\multicolumn{1}{l|}{O1-Pruner (32B)}     & 31.6 &  34.9 & 24.5 & 33.2 & \textbf{21.0} & 24.9 & \textbf{24.7} & 33.3 & 11.8 & 12.9 &  29.9 & 27.1  &         25.8 \\
\hline
\rowcolor{gray!20}
\multicolumn{14}{l}{\textit{\textbf{Ours}}}                                                                                                      \\
\hline
\multicolumn{1}{l|}{\ours (Qwen2.5-32B)} & 34.7 &  35.1 & 26.2 & 32.8 & 20.9 & 25.2 & 24.2 & 30.8 & 10.4 & 13.5 &  30.1 & 28.6  &          26.0 \\
\multicolumn{1}{l|}{\ours (QwQ-32B)}     & \textbf{35.2} &  \textbf{35.5} & \textbf{27.0} & \textbf{33.7} & 19.4 & \textbf{25.9} & 23.4 & 31.6 & 11.3 & \textbf{14.1} & \textbf{30.8} & \textbf{29.8}  & \textbf{26.5} \\
\hline
\end{tabular}
}
\caption{Retrieval performance (nDCG@10) on BRIGHT benchmark using \textbf{dense retriever (ReasonIR)} as the underlying retriever. All methods differ only in their reasoning strategy. Best scores per dataset are bolded.}
\label{tab:result_bright_reasonir}
\end{table*}

\subsubsection{RQ1: Does \ours Improve Retrieval Effectiveness?}

\paragraph{Overall Effectiveness}

To assess the retrieval effectiveness of \ours, we report the nDCG@10 scores of the final ranked documents on the BRIGHT benchmark which is a recent suite of tasks explicitly designed to evaluate retrieval performance under reasoning-intensive scenarios.
Evaluations on additional metrics such as MAP and Recall are provided in Appendix~\ref{appendix:result_bright_bm25_metric}.

\autoref{tab:result_bright_bm25} and \autoref{tab:result_bright_reasonir} compare our method against a variety of baselines, including standard CoT-based reranking approaches (Rank1, Rank-R1) and a compressed variant (O1-Pruner), under both sparse (BM25) and dense (DPR) retrieval settings.
Across all 12 domains, \ours consistently outperforms both standard and compressed CoT baselines.
Notably, both variants of \ours achieve the highest average nDCG@10 gain of \textbf{+5.4\%} on sparse retriever and \textbf{+2.1\%} on dense retriever, demonstrating that our structured reasoning framework remains highly effective in complex retrieval scenarios, regardless of the underlying LLM.

\begin{figure}[ht]
    \centering
    \includegraphics[width=1.0\columnwidth]{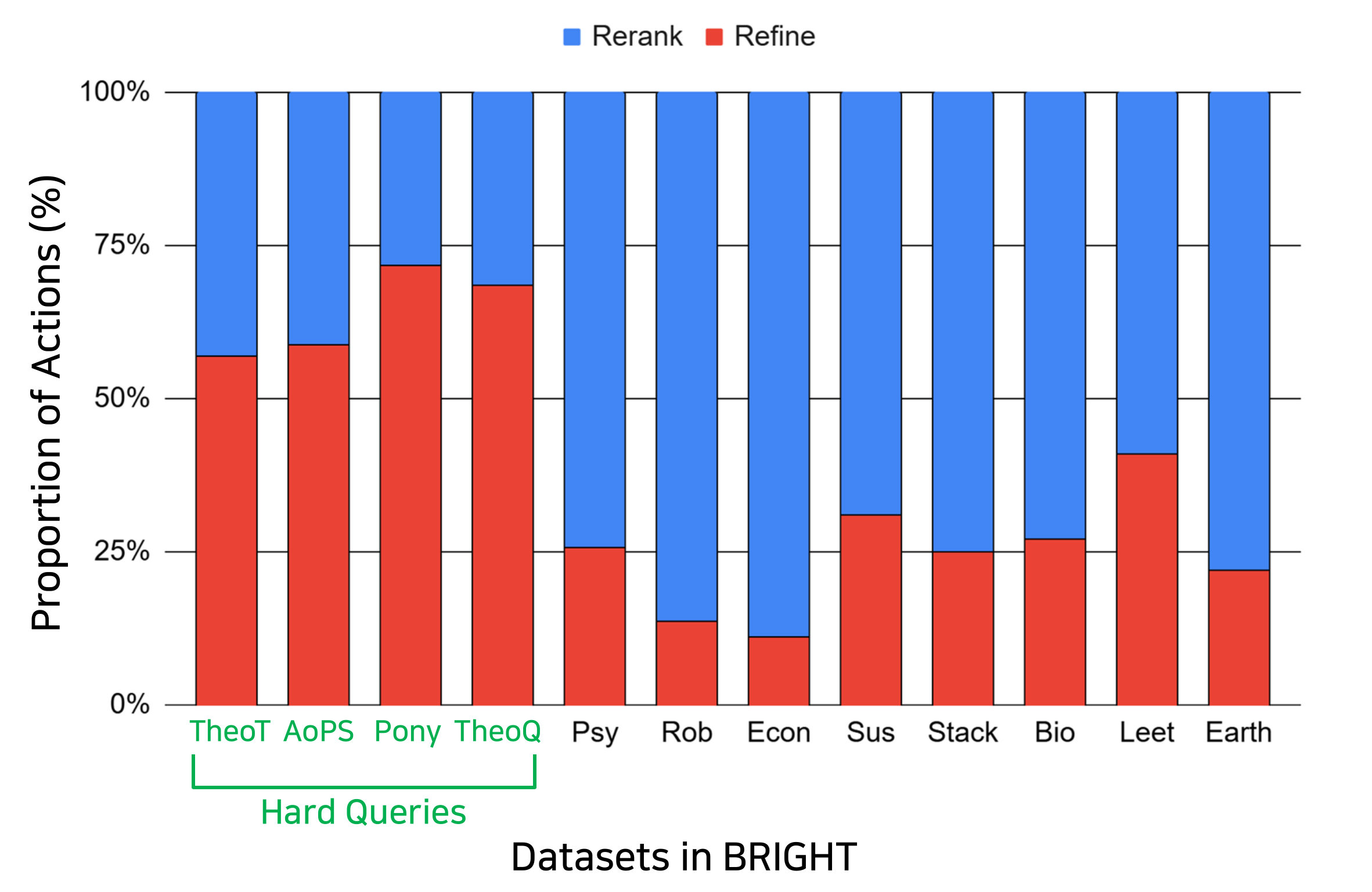}
    \caption{Distribution of reasoning actions selected by \ours on the BRIGHT benchmark. \textcolor{red}{Red} bars indicate the proportion of \textsc{Refine} actions, and \textcolor{blue}{blue} bars indicate \textsc{Rerank} actions.}
    \label{fig:action_ratio}
\end{figure}

\paragraph{Policy Effectiveness}

To further investigate the behavior of \ours, \autoref{fig:action_ratio} presents the distribution of selected actions across different datasets in the BRIGHT benchmark.
Each bar represents the relative frequency of the \textsc{Refine} and \textsc{Rerank} actions issued during reasoning.
The x-axis represents each dataset, sorted in ascending order of initial retriever performance, to examine how action distributions correlate with retrieval effectiveness.

We observe that \ours does not follow a fixed pattern or static heuristic.
Instead, action selection varies across tasks, reflecting the differing reasoning needs of each dataset.
In domains where the initial retrieval results are relatively informative (e.g., \textit{Bio}, \textit{Earth}, \textit{Econ}), the model predominantly uses \textsc{Rerank} to adjust document ordering while keeping the original query intact.
In contrast, in domains where the initial retrieval results are the lowest (\textcolor{ForestGreen}{green} subset) such as \textit{TheoT}, \textit{AoPS}, \textit{Pony}, \textit{TheoQ}, and they exhibit much higher proportions of \textsc{Refine} actions, up to \textbf{70\%} in some cases.

This pattern correlates strongly with underlying retrieval difficulty.
Faced with limited initial evidence, \ours adaptively runs multiple rounds of query refinement, attempting to surface better documents before stopping.
This behavior demonstrates that our action policy is not based on simple rule-based heuristics, but rather leverages contextual cues to make informed decisions.

Moreover, we conduct an ablation study comparing \ours with several fixed or naive strategies in Appendix~\ref{appendix:ablation_policy}, demonstrating the effectiveness of our prompt-based policy design.
Specifically, we compare our full policy with three simplified baselines:
one that uses only the \textsc{Refine} action, one that uses only \textsc{Rerank}, and one that selects actions uniformly at random.
\ours consistently achieves the best performance among these, indicating that our prompt-based policy design effectively balances refinement and reranking to support more accurate and efficient reasoning.

\paragraph{Intent Drift}

To assess whether iterative query refinement leads to intent drift, we conducted an automatic evaluation using the BRIGHT benchmark.
At each \textsc{Refine} step (up to 16 per query), we measured how well the rewritten query preserved the original intent.
We use GPT-4o-mini as the reference evaluator to compute alignment scores.
On average, across all datasets and refinement steps, intent alignment consistently remained above \textbf{0.9}, indicating that the refined queries retained strong fidelity to the user's original intent throughout the reasoning process.
More details are provided in Appendix~\ref{appendix:align_refine}.

\subsubsection{RQ2: Does \ours Enhance Token Efficiency?}

\begin{figure}[ht]
    \centering
    \includegraphics[width=1.0\columnwidth]{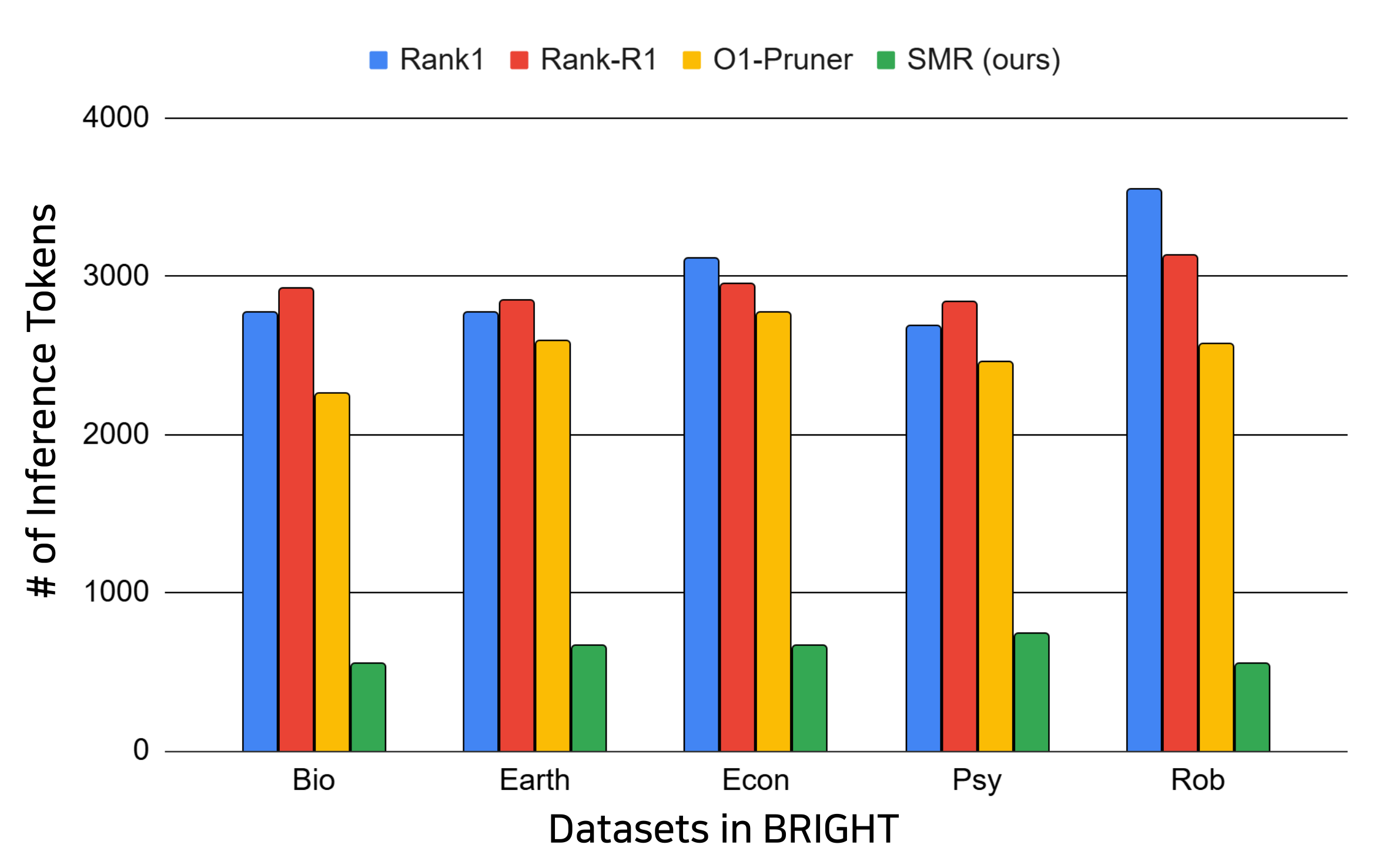}
    \caption{Inference token usage across five representative datasets in the BRIGHT benchmark. \ours (\textcolor{ForestGreen}{green} bars) achieves significantly lower token consumption than Rank1, Rank-R1, and O1-Pruner, while improving retrieval performance. Full results including all datasets are presented in Appendix~\ref{appendix:token_efficiency}.}
    \label{fig:token_efficiency}
\end{figure}

\paragraph{Token Efficiency}

To evaluate the token efficiency of our method, we compare the total number of inference tokens used by each system on the BRIGHT benchmark.
Notably, \ours executes both the policy decision and the \textsc{Refine}/\textsc{Rerank} steps using a single LLM via prompt-based interaction, without invoking any additional tools.
As a result, the total token count serves as a practical estimate of real-world inference latency.
Full results including all datasets are presented in Appendix~\ref{appendix:token_efficiency}.

As shown in \autoref{fig:token_efficiency}, \ours (\textcolor{ForestGreen}{green} bars) consistently consumes significantly fewer tokens across five representative datasets (Bio, Earth, Econ, Pay, Rob) compared to prior baselines including Rank1, Rank-R1, and O1-Pruner.
On average, \ours reduces inference token usage by \textbf{74.4\%}, outperforming even CoT compression methods such as O1-Pruner, which achieve only marginal reductions under \textbf{5\%}.

While these baselines often perform redundant reasoning steps, \ours minimizes such redundancy by proposing structured states and terminating reasoning early when convergence is detected.
This allows our method to achieve superior retrieval performance (see \autoref{tab:result_bright_bm25} and \autoref{tab:result_bright_reasonir}) while using substantially less computational resources.

\begin{figure}[ht]
    \centering
    \includegraphics[width=1.0\columnwidth]{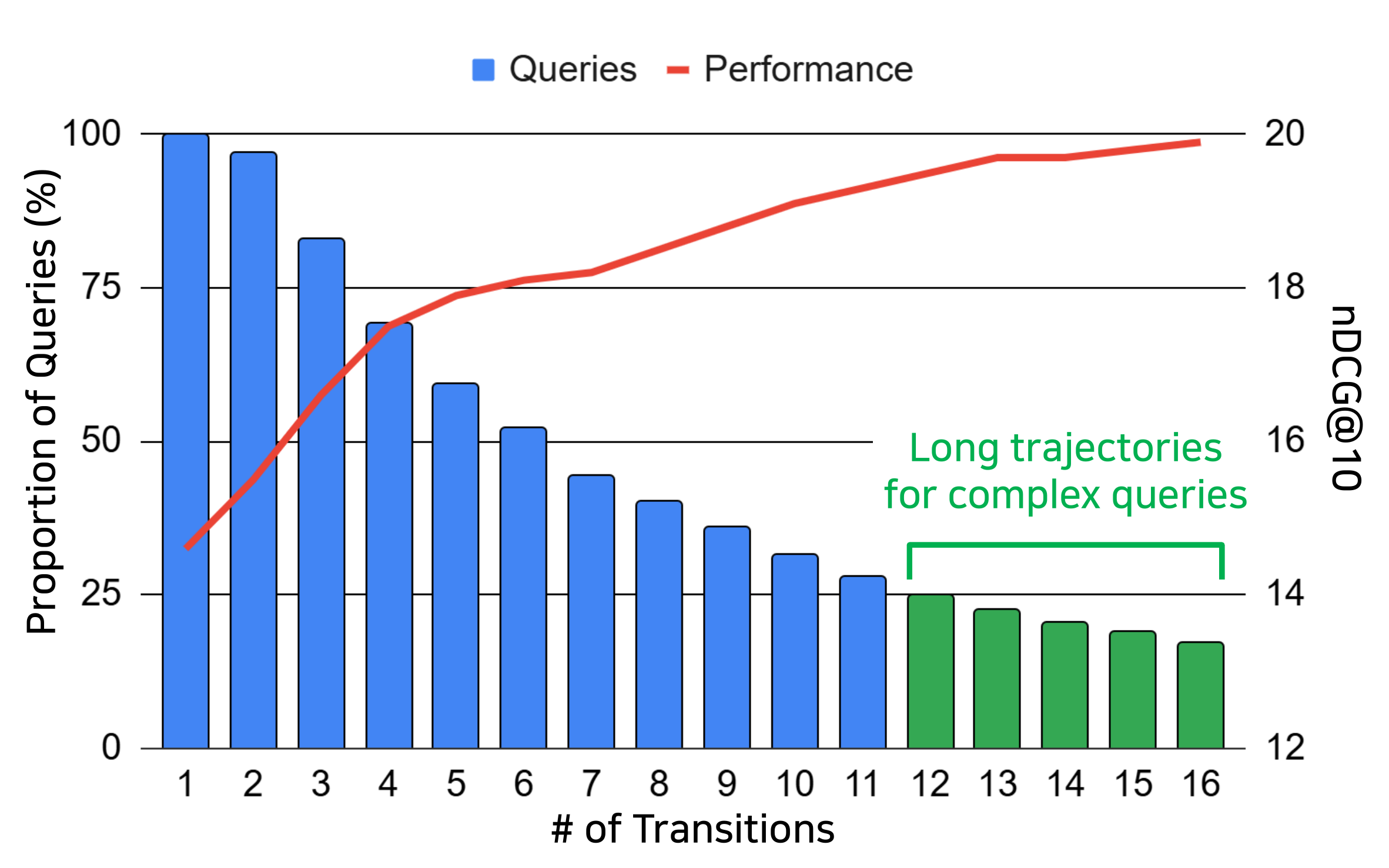}
    \caption{Transition statistics of \ours on the BRIGHT benchmark. The \textcolor{blue}{blue} bars indicate the number of reasoning steps across all queries, computed cumulatively per transition depth. The \textcolor{red}{red} curve shows the average retrieval performance (nDCG@10) at each step.}
    \label{fig:step_performance}
\end{figure}

\paragraph{Action Efficiency}
% \sw{how is it diff from token}
% To better understand the dynamics of our transition-based reasoning process, 
To evaluate how our policy model selects actions efficiently for queries of varying complexity,
we analyze the distribution of action lengths across all queries in the BRIGHT benchmark, as shown in \autoref{fig:step_performance}.
Each bar indicates how many queries reached each step in the reasoning actions, using a cumulative count, i.e., a query with a total length of 3 contributes to bins 1, 2, and 3.

We observe that \textbf{25\%} of queries terminate within \textbf{3 steps}, and \textbf{50\%} within \textbf{6 steps}, indicating that our system converges quickly.
This validates our early stopping mechanism and shows that many queries can be resolved efficiently without overthinking.

At the same time, \textbf{25\%} of queries (\textcolor{ForestGreen}{green} bar) exhibit \textbf{12 or more} actions, especially for complex or ambiguous inputs, which benefit from additional refinement and reranking.
This illustrates the flexibility of \ours which selectively allocates computation where needed.
Examples of such queries are provided in Appendix~\ref{appendix:example_transition}.
The examples show how \ours successfully navigates these hard queries, while the baselines fail to handle them.

Notably, the \textcolor{red}{red} curve in \autoref{fig:step_performance} shows the average retrieval performance (nDCG@10) after each transition step.
Retrieval performance improves steadily as reasoning progresses, demonstrating that our policy model incrementally updates the state with meaningful gains.
While we enforce a maximum of 16 steps due to computational constraints, the upward trend indicates that further improvements may be possible with extended actions, highlighting our capacity for continued enhancement when resources permit.

\subsubsection{RQ3: Is \ours Generally Applicable?}

\begin{table}[ht]
\centering
\scalebox{0.78}{
\def\arraystretch{1.3}
\begin{tabular}{lccc|c}
\hline
\multicolumn{1}{l|}{}                    & DBpedia & SciFact & FiQA &  \textbf{Avg} \\
\hline
\rowcolor{gray!20}
\multicolumn{5}{l}{\textit{\textbf{Retriever}}}                                     \\
\hline
\multicolumn{1}{l|}{GPL}                 &    36.1 &    66.5 & 32.9 &          45.2 \\
\hline
\rowcolor{gray!20}
\multicolumn{5}{l}{\textit{\textbf{CoT Reasoning}}}                                 \\
\hline
\multicolumn{1}{l|}{Rank1 (32B)}         &    38.2 &    70.2 & 35.8 &          48.1 \\
\multicolumn{1}{l|}{Rank-R1 (14B)}       &    38.9 &    71.6 & 37.1 &          49.2 \\
\hline
\rowcolor{gray!20}
\multicolumn{5}{l}{\textit{\textbf{Compressed CoT Reasoning}}}                      \\
\hline
\multicolumn{1}{l|}{O1-Pruner (32B)}     &    32.6 &    71.9 & 36.0 &          46.8 \\
\hline
\rowcolor{gray!20}
\multicolumn{5}{l}{\textit{\textbf{Ours}}}                                          \\
\hline
\multicolumn{1}{l|}{\ours (Qwen2.5-32B)} &    37.6 &    \textbf{73.1} & \textbf{38.7} & \textbf{49.8} \\
\multicolumn{1}{l|}{\ours (QwQ-32B)}     &    \textbf{39.5} &    71.2 & 36.5 &          49.1 \\
\hline
\end{tabular}
}
\caption{Retrieval performance (nDCG@10) on BEIR datasets. All methods use GPL~\cite{wang2021gpl} as the retriever, varying only in reasoning strategy. Best scores per dataset are bolded.}
\label{tab:result_beir}
\end{table}

\paragraph{Generalizability}

To assess the general applicability of \ours beyond reasoning-intensive scenarios, we evaluate its performance on the BEIR benchmark~\cite{thakur2021beir}, which covers a broad range of standard IR tasks without explicit reasoning demands.
\autoref{tab:result_beir} presents the nDCG@10 results on three BEIR datasets (DBpedia, SciFact, and FiQA).

\autoref{tab:result_beir} shows that \ours achieves the highest average performance among all methods, outperforming both standard CoT and a compressed reasoning approach.
This demonstrates that our structured reasoning approach is not specialized solely for reasoning-intensive scenarios, but instead transfers effectively to standard retrieval scenarios.

\section{Conclusion}
\label{sec:conclusion}
We presented \ours, a structured reasoning for retrieval by modeling reasoning as transitions over discrete states, defined as  queries and document rankings.
To mitigate overthinking, unlike prior methods such as CoT prompting or ReAct relying on token-level generation or tool calls, \ours takes an action-driven approach constraining reasoning to a set of IR-specific actions and grounding each step in a well-defined state.
 \ours offers a compact and controllable alternative to token-level reasoning with three key benefits:
(1) improved retrieval performance,
(2) reduced token usage through early stopping and redundancy checks, and
(3) robust generalization across retrievers and LLMs without task-specific tuning.
Experiments on the BRIGHT and BEIR benchmarks validate these advantages, with \ours outperforming strong baselines in both complex and standard IR settings.

\section{Limitation}
\label{sec:limitation}
While \ours offers a modular and token-efficient alternative to conventional CoT reasoning, our current state representation is limited to representing the query and the top-$k$ retrieved documents, and does not incorporate user interaction signals such as click-through rates or engagement metrics.
Incorporating such behavioral feedback could further enhance state fidelity and reasoning quality.
Moreover, while we restrict reasoning actions to \textsc{Refine}, \textsc{Rerank}, and \textsc{Stop}, the modular nature of our framework allows integration of additional operations such as domain-specific retrieval modules, document filters, or other tools without requiring architectural changes.

\bibliography{custom}

\appendix
\section{Appendix}
\label{sec:appendix}

\subsection{Details of Refine Action}
\label{appendix:detail_refine}
We avoid fully replacing $D_t$ with newly retrieved results.
Doing so risks erasing valuable signals from earlier reasoning steps, which can mislead the reasoning trajectory.
On the other hand, omitting retrieval altogether after a query refinement would be redundant, as the refined query is intended to improve document retrieval.
By integrating retrieval directly into the \textsc{Refine} action, we ensure consistency of reasoning and avoid unnecessary decomposition into separate actions.

\subsection{Policy Prompt}
\label{appendix:prompt_policy}

We provide the full prompt used to guide action selection in our system, as shown in \autoref{tab:prompt_policy}.
The prompt is designed to simulate the behavior of a decision-making agent responsible for managing a search process through discrete reasoning steps.

It describes three possible actions with their input and output formats, decision conditions, and justifications.
The prompt is instruction-based and free of task-specific training, enabling it to be applied across diverse retrieval scenarios without reconfiguration.

\begin{table*}[ht]
\rule{\textwidth}{0.4pt}
\begin{lstlisting}[basicstyle=\ttfamily\scriptsize, breaklines=true]
You are a highly intelligent artificial agent responsible for managing a search system. Your role is to either refine the given query or re-rank retrieved search results, thereby enhancing both recall and precision of the search. You can output exactly one of the following operations, after which another agent will execute it and return the results to you.

## Input Format
The input provided to you will have the following structure:

```
{
"query": "<current version of a query>",
"retrieved": [
    ("<docid>", "document contents"),
    ("<docid>", "document contents"),
    ...
]
}
```

### Decision policy (check in order):

1. Query Refinement
   Choose "refine query." if any of the following are met:
   - The query is ambiguous or generic
   - The retrieved search results are unsatisfactory
   - The query is short
   - Key domain terms are missing in the query

2. Reranking
   Only if the query already looks good and at least one retrieved document seems on topic.

3. Stop
   Only when you are certain that no further improvement is possible.


## Possible Outputs (select exactly one)

### Query Refinement
You may refine the query by rewriting it into a clear, specific, and formal version that is better suited for retrieving relevant information from a list of passages. Only return the document IDs (`docid`) in the `reranked` list. Do not include document contents. Output format:

```
{
"action": "refine query",
"refined_query": "<refined version of a query>",
"reason": "<reason for this action>"
}
```

### Re-ranking
You may reorder the retrieved documents (do not remove non-relevant ones). The results should be sorted in descending order of relevance. Output format:

```
{
"action": "re-rank",
"reranked": ["<docid>", "<docid>", ...],
"reason": "<reason for this action>"
}
```

### Stop
You may stop this iteration when the results are satisfactory. Output format:

```
{
"action": "stop"
}
```
\end{lstlisting}
\rule{\textwidth}{0.4pt}
\caption{Full prompt used in our system.}
\label{tab:prompt_policy}
\end{table*}

\subsection{Details of Policy}
\label{appendix:detail_policy}
LLM behavior during decoding is sensitive to temperature settings.
A low temperature (e.g., $T=0$) yields deterministic outputs, promoting stability, while higher temperatures increase diversity by allowing exploration of alternative generations.

We use the zero temperature to ensure consistency at default.
However, if the model fails to produce a valid output, such as malformed JSON or empty responses, we incrementally raise the temperature by 0.1 and retry.
This strategy expands the model’s output space just enough to recover from failure without sacrificing control.

Importantly, our reasoning framework is designed to tolerate such exploratory decoding.
Because each action operates within a strict interface and is validated at each step, the system remains robust even under increased temperature.
This allows us to combine deterministic defaults with controlled exploration for improved reliability.

\subsection{Implementation Details}
\label{appendix:implementation_details}
Rank1-32B is derived from Qwen2.5-32B\footnotemark[1], and Rank-R1-14B is derived from Qwen2.5-14B\footnotemark[2].
Since the 32B version is not available in Rank-R1, we experiment with the 14B model.
Since O1-Pruner is not originally designed for retrieval, we apply our prompt to perform query rewriting and document reranking using the model. O1-Pruner-32B is derived from QwQ-32B-preview\footnotemark[3].
To ensure a fair comparison given the model origin, we experiment with both particular non-reasoning and reasoning LLMs (Qwen2.5-32B\footnotemark[1], QwQ-32B\footnotemark[4]).

The token usage is computed by summing all generated output tokens across reasoning steps.
Input tokens are excluded to isolate the generation cost.
All experiments are run on a single NVIDIA A6000 GPU.
We set batch size, LLM temperature, top-$k$ retrieval, and the maximum number of reasoning steps for \ours as follows:
\texttt{batch\_size} = 8, \texttt{temperature} = 0.0, \texttt{k} = 10, \texttt{max\_steps} = 16.
These hyperparameters are held constant across all datasets and models unless otherwise stated.

\footnotetext[1]{\href{https://huggingface.co/Qwen/Qwen2.5-32B-Instruct}{Qwen/Qwen2.5-32B-Instruct}}
\footnotetext[2]{\href{https://huggingface.co/Qwen/Qwen2.5-14B-Instruct}{Qwen/Qwen2.5-14B-Instruct}}
\footnotetext[3]{\href{https://huggingface.co/Qwen/QwQ-32B-Preview}{Qwen/QwQ-32B-Preview}}
\footnotetext[4]{\href{https://huggingface.co/Qwen/QwQ-32B}{Qwen/QwQ-32B}}

\subsection{Evaluation on Additional Metrics}
\label{appendix:result_bright_bm25_metric}

To complement the main evaluation in terms of nDCG@10, we report additional metrics to further validate the robustness of our approach.
\autoref{tab:result_bright_bm25_map10}, and \autoref{tab:result_bright_bm25_recall10} show results on MAP@10 and Recall@10, respectively, evaluated on the BRIGHT benchmark using BM25 as the base retriever.
Across all metrics and datasets, \ours consistently achieves top performance compared to CoT and compressed CoT baselines.

A potential concern with our early stopping mechanism is that it may prematurely terminate the reasoning process, potentially reducing the recall of retrieved documents.
However, our results show that \ours consistently achieves higher recall than all baselines.
This indicates that the risk of under-retrieval is either negligible or outweighed by the benefits of our reasoning process.

\begin{table*}[t]
\centering
\scalebox{0.8}{
\def\arraystretch{1.3}
\begin{tabular}{lcccccccccccc|c}
\hline
\multicolumn{1}{l|}{}                    & Bio & Earth & Econ & Psy & Rob & Stack & Sus & Leet & Pony & AoPS & TheoQ & TheoT & \textbf{Avg} \\
\hline
\rowcolor{gray!20}
\multicolumn{14}{l}{\textit{\textbf{Retriever}}} \\
\hline
\multicolumn{1}{l|}{BM25}                & 13.1 & 18.1 & 8.9 & 7.0 & 8.5 & 13.4 & 10.0 & 19.7 & 1.6 & \textbf{6.2} & 10.4 & 4.9 & 10.2 \\
\hline
\rowcolor{gray!20}
\multicolumn{14}{l}{\textit{\textbf{CoT Reasoning}}} \\
\hline
\multicolumn{1}{l|}{Rank1 (32B)}         & 16.1 & 23.0 & 7.9 & 11.3 & 10.5 & 12.3 & 16.2 & 17.6 & \textbf{2.2} & 3.1 & 10.4 & 8.6 & 11.6 \\
\multicolumn{1}{l|}{Rank-R1 (14B)}       & 17.7 & 25.0 & 10.3 & 10.7 & 14.1 & 13.1 & 16.1 & 20.2 & 2.1 & 4.0 & 10.8 & 7.7 & 12.7 \\
\hline
\rowcolor{gray!20}
\multicolumn{14}{l}{\textit{\textbf{Compressed CoT Reasoning}}} \\
\hline
\multicolumn{1}{l|}{O1-Pruner (32B)}     & 17.2 & 22.6 & 10.7 & 12.9 & 12.7 & 14.9 & 15.5 & 19.5 & 1.8 & 3.7 & 8.9 & 5.6 & 12.2 \\
\hline
\rowcolor{gray!20}
\multicolumn{14}{l}{\textit{\textbf{Ours}}} \\
\hline
\multicolumn{1}{l|}{\ours (Qwen2.5-32B)} & 21.8 & 25.6 & \textbf{13.9} & 14.5 & \textbf{16.1} & 14.4 & 14.3 & 17.8 & 1.1 & 3.6 & \textbf{14.6} & \textbf{16.5} & 14.5 \\
\multicolumn{1}{l|}{\ours (QwQ-32B)}     & \textbf{22.0} & \textbf{25.7} & 13.6 & \textbf{15.4} & 15.0 & \textbf{15.5} & \textbf{16.6} & \textbf{21.0} & 1.6 & 3.1 & 14.1 & 15.2 & \textbf{14.9} \\
\hline
\end{tabular}
}
\caption{Retrieval performance (MAP@10) on BRIGHT benchmark using \textbf{sparse retriever (BM25)} as the underlying retriever. All methods differ only in their reasoning strategy. Best scores per dataset are bolded.}
\label{tab:result_bright_bm25_map10}
\end{table*}
\begin{table*}[t]
\centering
\scalebox{0.78}{
\def\arraystretch{1.3}
\begin{tabular}{lcccccccccccc|c}
\hline
\multicolumn{1}{l|}{}                    & Bio & Earth & Econ & Psy & Rob & Stack & Sus & Leet & Pony & AoPS & TheoQ & TheoT & \textbf{Avg} \\
\hline
\rowcolor{gray!20}
\multicolumn{14}{l}{\textit{\textbf{Retriever}}} \\
\hline
\multicolumn{1}{l|}{BM25}                & 21.7 & 31.9 & 16.8 & 15.6 & 19.6 & 21.3 & 21.3 & \textbf{29.5} & \textbf{4.1} & 6.0 & 8.6 & 9.2 & 17.1
 \\
\hline
\rowcolor{gray!20}
\multicolumn{14}{l}{\textit{\textbf{CoT Reasoning}}} \\
\hline
\multicolumn{1}{l|}{Rank1 (32B)}         & 21.7 & 31.9 & 16.8 & 15.6 & 19.6 & 21.3 & 21.3 & \textbf{29.5} & \textbf{4.1} & 6.0 & 8.6 & 9.2 & 17.1
 \\
\multicolumn{1}{l|}{Rank-R1 (14B)}       & 21.7 & 31.9 & 16.8 & 15.6 & 19.6 & 21.3 & 21.3 & \textbf{29.5} & \textbf{4.1} & 6.0 & 8.6 & 9.2 & 17.1
 \\
\hline
\rowcolor{gray!20}
\multicolumn{14}{l}{\textit{\textbf{Compressed CoT Reasoning}}} \\
\hline
\multicolumn{1}{l|}{O1-Pruner (32B)}     & 23.8 & \textbf{34.4} & \textbf{20.2} & 19.8 & 19.6 & 21.3 & 21.5 & \textbf{29.5} & 2.3 & 6.0 & 12.9 & 10.3 & 18.5 \\
\hline
\rowcolor{gray!20}
\multicolumn{14}{l}{\textit{\textbf{Ours}}} \\
\hline
\multicolumn{1}{l|}{\ours (Qwen2.5-32B)} & 25.3 & 33.0 & 19.5 & \textbf{21.7} & \textbf{21.4} & \textbf{24.6} & 21.6 & 26.6 & 3.6 & 6.1 & \textbf{20.9} & \textbf{24.8} & \textbf{20.8} \\
\multicolumn{1}{l|}{\ours (QwQ-32B)}     & \textbf{25.7} & 31.9 & 18.6 & 21.5 & 20.5 & 24.3 & \textbf{23.9} & \textbf{29.5} & 4.0 & \textbf{6.3} & 18.7 & 19.0 & 20.3 \\
\hline
\end{tabular}
}
\caption{Retrieval performance (Recall@10) on BRIGHT benchmark using \textbf{sparse retriever (BM25)} as the underlying retriever. All methods differ only in their reasoning strategy. Best scores per dataset are bolded.}
\label{tab:result_bright_bm25_recall10}
\end{table*}

\subsection{Ablation Study for Action Selection Policy}
\label{appendix:ablation_policy}

\begin{table*}[t]
\centering
\scalebox{0.78}{
\def\arraystretch{1.3}
\begin{tabular}{l|ccccc|c}
\hline
 & Bio & Earth & Econ & Psy & Rob & \textbf{Avg} \\
\hline
\ours{} (Qwen2.5-32B)               & \textbf{28.7} & \textbf{34.9} & \textbf{20.4} & 20.7 & \textbf{20.9} & \textbf{25.1} \\
\quad [\textsc{Rerank} or \textsc{Stop}]      & 23.9 & 33.6 & 17.7 & 15.8 & 19.1 & 22.0 \\
\quad [\textsc{Refine} or \textsc{Stop}]      & 26.8 & 29.5 & 14.4 & \textbf{20.9} & 12.4 & 20.8 \\
\quad [Randomly Selecting Action]                & 22.6 & 28.8 & 16.4 & 16.1 & 15.0 & 19.8 \\
\hline
\end{tabular}
}
\caption{Ablation study of policy choices for \ours{} (Qwen2.5-32B), measured by nDCG@10 across the first five domains.}
\label{tab:ablation_qwen2.5}
\end{table*}

We conduct an ablation study comparing \ours with several fixed or naive strategies in \autoref{tab:ablation_qwen2.5}, demonstrating the effectiveness of our prompt-based policy design.
Specifically, we compare \ours{} with three simplified variants: one that removes the \textsc{Refine} action ([\textsc{Rerank} or \textsc{Stop}]), one that removes the \textsc{Rerank} action ([\textsc{Refine} or \textsc{Stop}]), and one that randomly selects an action at each step.
We do not ablate \textsc{Stop} since it is essential for termination, but instead test whether random policy decisions degrade performance.
\ours consistently achieves the best performance among these, indicating that our prompt-based policy design effectively balances refinement and reranking to support more accurate and efficient reasoning.

\subsection{Evaluation of Query Intent Preservation During Refinement}
\label{appendix:align_refine}
\begin{table*}[ht]
\rule{\textwidth}{0.4pt}
\begin{lstlisting}[basicstyle=\ttfamily\scriptsize, breaklines=true]
Please compare the second query aligns with the any intent of the first query and give a score from 0 to 1.
The first query is the original query, and the second query is the modified query.
The score should be based on how similar the two queries are. You must output the floating number score only.
The first query is: "{query_original}".
The second query is: "{query}".
\end{lstlisting}
\rule{\textwidth}{0.4pt}
\caption{The prompt used to get alignment score between the original and refined queries.}
\label{tab:prompt_align_refine}
\end{table*}

\begin{table*}[ht]
\centering
\scalebox{0.75}{
\def\arraystretch{1.3}
\begin{tabular}{c|cccccccccccccccc}
\hline
 & 1 & 2 & 3 & 4 & 5 & 6 & 7 & 8 & 9 & 10 & 11 & 12 & 13 & 14 & 15 & 16 \\
\hline
Alignment Score & 0.98 & 0.97 & 0.95 & 0.95 & 0.95 & 0.94 & 0.94 & 0.93 & 0.93 & 0.93 & 0.92 & 0.92 & 0.92 & 0.92 & 0.91 & 0.91 \\
\hline
\end{tabular}
}
\caption{Average alignment scores across all datasets within BRIGHT.}
\label{tab:align_refine}
\end{table*}

To address concerns about potential query intent drift during iterative refinement, we conducted an automatic evaluation to quantify how well each rewritten query preserves the user’s original intent.
Specifically, we evaluated queries generated during each reasoning step of \textsc{Refine} using the BRIGHT benchmark.

We employed GPT-4o-mini as a reference evaluator.
For each reasoning step the model was shown the original query, the current rewritten query, and the retrieved documents used to generate.
The model was instructed to assign a continuous score between 0 and 1, where 1 indicates perfect preservation of the original intent and 0 indicates complete divergence.
The exact prompt used for this evaluation is included in \autoref{tab:prompt_align_refine}.

\autoref{tab:align_refine} shows the average alignment score across all datasets within BRIGHT and over the full trajectory of refinement steps (up to 16 per query).
The result shows that the average intent alignment score remained above \textbf{0.9} at every step.
This suggests that, despite iterative rewriting, the refined queries remain closely aligned with the initial user intent.
We attribute this stability to the fact that each refinement is grounded in both the immediately preceding query and its associated retrieval context, which anchors the model’s generation.
This evaluation provides quantitative evidence that our modular \textsc{Refine} step does not introduce meaningful semantic drift, supporting the claim that our reasoning procedure maintains alignment with user goals throughout the refinement trajectory.

\subsection{Token Efficiency}
\label{appendix:token_efficiency}

\ours achieves substantial reductions in inference token usage compared to all baseline methods across the full BRIGHT benchmark.
\autoref{tab:token_bright_bm25} shows the total number of tokens consumed by each method across individual tasks, using BM25 as the underlying retriever.

Our method consistently outperforms Rank1, Rank-R1, and O1-Pruner by large margins.
While baselines often produce long and redundant reasoning chains, \ours avoids such inefficiency through its structured state representation and early stopping mechanism.
This result further supports our claim that \ours can deliver better performance with fewer computational resources.

\begin{table*}[t]
\centering
\scalebox{0.77}{
\def\arraystretch{1.3}
\begin{tabular}{lcccccccccccc|c}
\hline
\multicolumn{1}{l|}{}                    &  Bio & Earth & Econ &  Psy &  Rob & Stack &  Sus & Leet & Pony & AoPS & TheoQ & TheoT &  \textbf{Avg} \\
\hline
\rowcolor{gray!20}
\multicolumn{14}{l}{\textit{\textbf{CoT Reasoning}}}                                                                                             \\
\hline
\multicolumn{1}{l|}{Rank1 (32B)}         & 2,772 & 2,775 & 3,115 & 2,688 & 3,548 & 3,554 & 2,691 & 3,248 & 2,570 & 3,390 & 3,135 & 3,281 & 3,064 \\
\multicolumn{1}{l|}{Rank-R1 (14B)}       & 2,924 & 2,842 & 2,951 & 2,835 & 3,130 & 3,065 & 2,895 & 3,020 & 3,383 & 2,674 & 2,680 & 3,379 & 2,982 \\
\hline
\rowcolor{gray!20}
\multicolumn{14}{l}{\textit{\textbf{Compressed CoT Reasoning}}}                                                                                  \\
\hline
\multicolumn{1}{l|}{O1-Pruner (32B)}     & 2,259 & 2,595 & 2,771 & 2,457 & 2,568 & 2,818 & 2,492 & 6,187 & 2,870 & 4,126 & 2,520 & 2,375 & 3,003 \\
\hline
\rowcolor{gray!20}
\multicolumn{14}{l}{\textit{\textbf{Ours}}}                                                                                                      \\
\hline
\multicolumn{1}{l|}{\ours (Qwen2.5-32B)} & \textbf{551} & \textbf{666} & \textbf{663} & \textbf{741} & \textbf{552} & \textbf{794} & \textbf{598} & 982 & \textbf{780} & 1,063 & 914 & 996 & \textbf{775} \\
\multicolumn{1}{l|}{\ours (QwQ-32B)}     & 1,891 & 1,167 & 1,594 & 2,114 & 1,465 & 1,144 & 2,034 & \textbf{672} & 1,094 & \textbf{897} & \textbf{811} & \textbf{805} & 1,307 \\
\hline
\end{tabular}
}
\caption{Total number of tokens generated during inference on the BRIGHT benchmark. The lowest token counts, indicating higher inference efficiency, are marked in bold.}
\label{tab:token_bright_bm25}
\end{table*}

\subsection{Examples of Hard Queries}
\label{appendix:example_transition}
\begin{table*}[ht!]
\scriptsize
\centering
\setlength{\tabcolsep}{4pt}
\renewcommand{\arraystretch}{1.3}
\begin{tabularx}{\textwidth}{%
    p{1cm}
    p{2cm}
    >{\raggedright\arraybackslash}X
}
\toprule
\textbf{Iter} & \textbf{Field} & \textbf{Content} \\ 
\midrule
\multirow{2}{*}{Iter 0}
  & Query ($q_{0}$)     & Imagine you're playing a game where you have a toy cannon that shoots small metal balls. You decide to aim at a lightweight door that can swing open on its hinges. The door is 1 meter wide and weighs 15 kilograms. You shoot a metal ball weighing 10 grams at a speed of 400 meters per second straight at the middle of the door, and the ball sticks to it. What is the new angular speed of the door after the ball sticks to it? (Unit: rad/s)
    \\ \cmidrule(lr){2-3}
  & Document ($D_{0}$)  &  $[d_1]$, $[d_2]$(about chess), $[d_3]$(about Einstein’s equation) …, $[d_{k}]$
    \\ 
\midrule
\multirow{4}{*}{Iter 1}
  & Action              &  \textsc{Rerank}
    \\ \cmidrule(lr){2-3}
  & Query ($q_{1}$)     &  Imagine you're playing a game where you have a toy cannon that shoots small metal balls. You decide to aim at a lightweight door that can swing open on its hinges. The door is 1 meter wide and weighs 15 kilograms. You shoot a metal ball weighing 10 grams at a speed of 400 meters per second straight at the middle of the door, and the ball sticks to it. What is the new angular speed of the door after the ball sticks to it? (Unit: rad/s)
    \\ \cmidrule(lr){2-3}
  & Document ($D_{1}$)  & $[d_1]$, $[d_{k}]$, ... $[d_2]$(about chess), $[d_3]$(about Einstein’s equation)
    \\ \cmidrule(lr){2-3}
  & Reason              &  \makecell[tl]{
      $[d_1]$: ... \\ 
      $[d_2]$: Okay, let's see. The user is asking whether the passage is relevant to calculating the angular speed ... \\ 
      \textbf{There's no overlap in subject matter here.} ...\\
      Even though both involve calculations, \textbf{the context and required knowledge are completely different.} ...\\
      \textbf{The chess passage is about combinatorics and game theory, while the query is a physics problem.} ...\\
     \textbf{They cover completely different topics with no overlap in relevant concepts.} ...\\
      Therefore, the passage isn't relevant to the query. ... \\
      $[d_3]$: ... \\
      ... \\
      \\
      \textit{[number of tokens: 2542]}
    }
    \\ 
\bottomrule
\end{tabularx}
\caption{A qualitative example of redundant reasoning in Rank1, where the model repeats semantically equivalent steps.}
\label{tab:example_rank1}
\end{table*}

\begin{table*}[ht!]
\scriptsize
\centering
\setlength{\tabcolsep}{4pt}
\renewcommand{\arraystretch}{1.3}
\begin{tabularx}{\textwidth}{
    p{1cm}
    p{2cm}
    >{\raggedright\arraybackslash}X
}
\toprule
\textbf{Iter} & \textbf{Field} & \textbf{Content} \\ 
\midrule
\multirow{2}{*}{Iter 0}
  & Query ($q_{0}$)     & Imagine you're playing a game where you have a toy cannon that shoots small metal balls. You decide to aim at a lightweight door that can swing open on its hinges. The door is 1 meter wide and weighs 15 kilograms. You shoot a metal ball weighing 10 grams at a speed of 400 meters per second straight at the middle of the door, and the ball sticks to it. What is the new angular speed of the door after the ball sticks to it? (Unit: rad/s)
    \\ \cmidrule(lr){2-3}
  & Document ($D_{0}$)  &  $[d_1]$, $[d_2]$(about chess), $[d_3]$(about Einstein’s equation)\dots, $[d_{k}]$
    \\ 
\midrule
\multirow{3}{*}{Iter 1}
  & Action              &  \textsc{Refine}
    \\ \cmidrule(lr){2-3}
  & Query ($q_{1}$)     &  Calculate angular speed after a collision using conservation of angular momentum
    \\ \cmidrule(lr){2-3}
  & Reason              &  \makecell[tl]{
      Alright, I’ve got this problem here about a toy cannon shooting a metal ball at a door,\\ 
      and I need to find the new angular speed of the door after the ball sticks to it. \dots\\
      It’s about angular speed after a collision. \dots\\
      Specifically, it’s about conservation of angular momentum. \dots\\
      Again, I need the angular speed once the ball sticks to the door. \dots\\
      Again, conservation of angular momentum governs the result. \dots\\
      \textbf{Perhaps phrased as “calculate angular speed after a collision using conservation of angular momentum.”} \dots \\
      \\
      \textit{[number of tokens: 1153]}
    }
    \\ 
\midrule
\multirow{4}{*}{Iter 2}
  & Action              &  \textsc{Rerank}
    \\ \cmidrule(lr){2-3}
  & Query ($q_{2}$)     &  Calculate angular speed after a collision using conservation of angular momentum
    \\ \cmidrule(lr){2-3}
  & Document ($D_{1}$)  &  $[d_1]$, $[d_{k}]$, \dots, $[d_2]$(about chess), $[d_3]$(about Einstein’s equation)
    \\ \cmidrule(lr){2-3}
  & Reason              &  \makecell[tl]{%
      $[d_1]$: Relevant ...\\ 
      $[d_2]$: Not relevant ...\\
      $[d_3]$: Not relevant ...\\
      ... \\
      \\
      \textit{[number of tokens: 992]} \\
      \textit{[total number of tokens: 2143]}
    }
    \\ 
\bottomrule
\end{tabularx}
\caption{A qualitative example of misguided reasoning in O1-Pruner, where compression produces a concise but semantically misaligned query.}
\label{tab:example_o1_pruner}
\end{table*}
\begin{table*}[ht!]
\scriptsize
\centering
\setlength{\tabcolsep}{4pt}
\renewcommand{\arraystretch}{1.3}
\begin{tabularx}{\textwidth}{
    p{1cm}
    p{2cm}
    >{\raggedright\arraybackslash}X
}
\toprule
\textbf{Iter} & \textbf{Field} & \textbf{Content} \\ 
\midrule
\multirow{2}{*}{Iter 0}
  & Query ($q_{0}$)     & Imagine you're playing a game where you have a toy cannon that shoots small metal balls. You decide to aim at a lightweight door that can swing open on its hinges. The door is 1 meter wide and weighs 15 kilograms. You shoot a metal ball weighing 10 grams at a speed of 400 meters per second straight at the middle of the door, and the ball sticks to it. What is the new angular speed of the door after the ball sticks to it? (Unit: rad/s) 
    \\ \cmidrule(lr){2-3}
  & Document ($D_{0}$)  &  $[d_1]$, $[d_2]$(about chess), $[d_3]$(about Einstein's equation) …, $[d_{k}]$ 
    \\ 
\midrule
\multirow{3}{*}{Iter 1}
  & Action              &  \textsc{Refine} 
    \\ \cmidrule(lr){2-3}
  & Query ($q_{1}$)     &  How to calculate angular speed (rad/s) of a door after collision with metal ball using conservation of angular momentum? Door width=1m, mass=15kg; ball:mass=0.01kg, velocity=400 m/s hits center and sticks. 
    \\ \cmidrule(lr){2-3}
  & Reason              &  \makecell[tl]{
      The original query is missing key physics concepts like \textbf{'conservation of angular momentum' which are essential for} \\ 
      \textbf{solving the problem.} The retrieved documents include unrelated topics (e.g., Einstein's equation, chess), so refining to focus \\
      on angular momentum will improve precision. \\
      \\
      \textit{[number of tokens: 47]}
    }
    \\ 
\midrule
\multicolumn{1}{c}{\LARGE{…}} & & \\
\midrule
\multirow{4}{*}{Iter 5}
  & Action              &  \textsc{Rerank} 
    \\ \cmidrule(lr){2-3}
  & Query ($q_{5}$)     &  How to calculate angular speed (rad/s) of a door after collision with metal ball using conservation of angular momentum? Door width=1m, mass=15kg; ball:mass=0.01kg, velocity=400 m/s hits center and sticks. 
    \\ \cmidrule(lr){2-3}
  & Document ($D_{5}$)  &  $[d_1]$, $[d_{k}]$, …, $[d_2]$(about chess), $[d_3]$(about Einstein’s equation) 
    \\ \cmidrule(lr){2-3}
  & Reason              &  \makecell[tl]{
      The retrieved document ($[d_1]$) include foundational content on \textbf{angular momentum, units of mass and length} relevant to  \\
      the problem's parameters. \\
      Document $[d_{k}]$ discusses \textbf{angular velocity as a vector quantity which is essential for calculating direction-dependent} \\
      \textbf{quantities like angular speed.} \\ \textbf{The other documents either discuss unrelated topics} such as chess move counts or are less directly related.
      … \\
      \\
      \textit{[number of tokens: 67]}
    }
    \\ 
\midrule
\multirow{1}{*}{Iter 6}
  & Action              &  \makecell[tl]{
      \textsc{Stop} \\
      \\
      \textit{[total number of tokens: 252]}
    }
    \\
\bottomrule
\end{tabularx}
\caption{A qualitative example of SMR (QWQ-32B), demonstrating efficient reasoning in IR.}
\label{tab:example_smr}
\end{table*}

To concretely illustrate how \ours addresses the two central challenges identified in Section~\ref{sec:intro}, we present an example drawn from the BRIGHT benchmark involving a complex query.
This case also corresponds to the \textbf{hard queries} analyzed in \autoref{fig:step_performance}, where longer reasoning is needed to recover relevant evidence.

The query asks for the angular speed of a door after a perfectly inelastic collision with a metal ball.
Although the problem is detailed and well-structured, it includes distractor terms such as \textit{game}, \textit{toy}, and \textit{door}, which cause initial retrieval to return irrelevant content, including passages on chess and probability.
This scenario serves as a testbed for evaluating reasoning robustness under lexical noise and semantic mismatch.

\paragraph{Rank1 (Redundant Reasoning)}
Rank1 (\autoref{tab:example_rank1}) generates token-level CoT traces, and the model enters a redundant loop.
It repeatedly justifies why the chess passages are irrelevant, consuming over 2,500 tokens without progressing the retrieval state.
Although its judgment is eventually correct, the system lacks a mechanism to recognize semantic equivalence between states and continues reasoning far beyond necessity.
This exemplifies the first challenge, \textbf{redundant trajectories}, where the model revisits similar states without gaining new information.

\paragraph{O1-Pruner (Misguided Reasoning)}
O1-Pruner (\autoref{tab:example_o1_pruner}) aggressively compresses the reasoning trajectory using reinforcement learning.
It quickly rewrites the query into a concise fragment:
``calculate angular speed after a collision using conservation of angular momentum.'' While technically correct, this abstraction removes too much context from the original problem,
leading the retriever to return generic physics content that fails to address the door scenario.
This case illustrates the second challenge, \textbf{misguided reasoning}, where brevity is achieved at the cost of alignment with the intent of the user.

\paragraph{\ours (Ours)}
By contrast, \ours (\autoref{tab:example_smr}) successfully navigates this scenario using structured state transitions.
It first refines the query to include all necessary physics parameters, and then reranks documents to surface the most relevant passages.
Each updated state is checked for semantic equivalence to prevent drift.
Because actions are discrete and grounded in context, \ours avoids misalignment and redundant loops, achieving efficient reasoning in IR.

\subsection{Usage of AI Assistants}
ChatGPT was employed to enhance the clarity and grammatical accuracy of the text, offering suggestions for sentence rephrasing and correction of grammatical errors.

\end{document}